\documentclass[a4paper]{article}

\usepackage{INTERSPEECH2022}
\usepackage{multirow}
\usepackage[table,x11names,dvipsnames]{xcolor}
\usepackage{graphicx}
\usepackage{url}
\usepackage{hyperref}
\usepackage{cite}
\usepackage[normalem]{ulem}
\usepackage{hhline}
\DeclareRobustCommand{\sbseries}{\fontseries{sb}\selectfont}
\DeclareTextFontCommand{\textsb}{\sbseries}

\title{Hybrid Handcrafted and Learnable Audio Representation for\\Analysis of Speech Under Cognitive and Physical Load}

\name{Gasser Elbanna$^{1,2*}$\thanks{$^*$GE performed this work as an intern at Logitech.}, Alice Biryukov$^{1,2}$, Neil Scheidwasser-Clow$^{1}$, Lara Orlandic$^{1}$, Pablo Mainar$^2$, Mikolaj Kegler$^3$, Pierre Beckmann$^4$, Milos Cernak$^2$}
\address{
  $^1$École Polytechnique Fédérale de Lausanne (EPFL), Lausanne, Switzerland\\
  $^2$Logitech Europe S.A., Lausanne, Switzerland\\
  $^3$Imperial College London, London, United Kingdom\\
  $^4$Université de Lausanne, Lausanne, Switzerland}
  
\email{gasser.elbanna@epfl.ch, milos.cernak@ieee.org}

\begin{document}
\maketitle

\begin{abstract}

As a neurophysiological response to threat or adverse conditions, stress can affect cognition, emotion and behaviour with potentially 
detrimental effects on health in the case of sustained exposure. 
Since the affective content of speech is inherently modulated by an individual's physical and mental state, a substantial body of research has been devoted to the study of paralinguistic correlates of stress-inducing task load.
Historically, voice stress analysis has been conducted using conventional digital signal processing (DSP) techniques.
Despite the development of modern methods based on deep neural networks (DNNs), accurately detecting stress in speech remains difficult due to the wide variety of stressors and considerable variability in individual stress perception. To that end, we introduce a set of five datasets for task load detection in speech. The voice recordings were collected as either cognitive or physical stress was induced in the cohort of volunteers, with a cumulative number of more than a hundred speakers. We used the datasets to design and evaluate a novel self-supervised audio representation that leverages the effectiveness of handcrafted features (DSP-based) and the complexity of data-driven DNN representations. Notably, the proposed approach outperformed both extensive handcrafted feature sets and novel DNN-based audio representation learning approaches.

\end{abstract}


\section{Introduction}
\label{sec:intro}
    


The emergence of machine learning methods for the paralinguistic analysis of speech signals has been of considerable influence for healthcare-related applications. This includes both the detection of long-term traits such as speech pathology \cite{vasquez2018} and shorter-term mental states such as emotion, effort, and acute stress \cite{schuller2013interspeech, schuller2014interspeech, puyvelde2018}. For the latter, inter-individual differences in stress perception constitutes, among others, one of the major limitations of voice stress analysis (VSA) studies \cite{giddens2013vocal}.



\begin{figure}[t]
    \centering
    \includegraphics[trim={0 0.23cm 0 0.1cm}, clip, width=\linewidth]{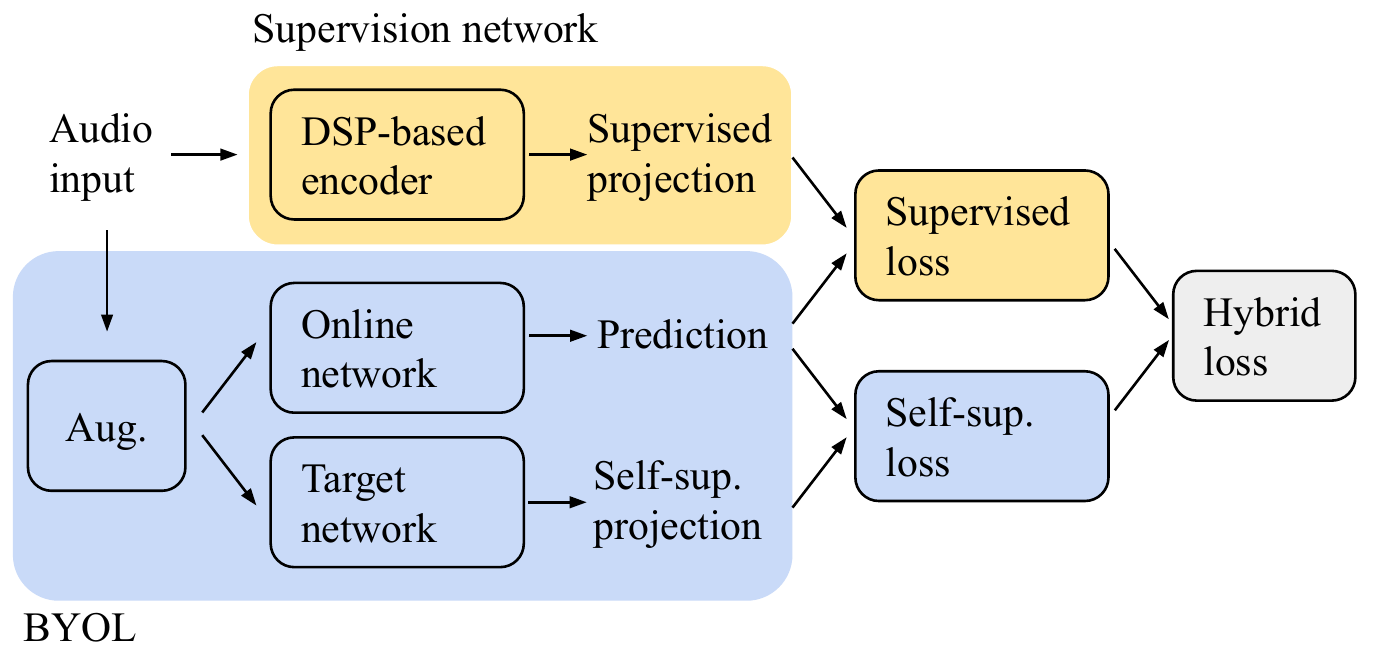}
    \caption{Hybrid audio representation learning: combining the self-supervised BYOL framework (blue) with supervision from DSP features (yellow).}\vspace{-8pt}
    \label{fig:diagram}
\end{figure}


Among the main sources of acute stress \cite{puyvelde2018}, effort-induced stress can be elicited by increased physical or cognitive demands to complete the task (i.e., physical and cognitive load, respectively). Both behavioural states have been the subject of numerous paralinguistic studies \cite{baker2008ventilation, johannes2007, trouvain2015, godin2008analysis, rothkrantz2004voice}. On the one hand, physical load is known to create notable differences in speech production, especially due to competition between breathing processes for speech generation and increased oxygen consumption to meet metabolic needs \cite{baker2008ventilation}. Regarding voice-related low-level descriptors (LLDs), several studies have reported increases in mean fundamental frequency after intensive physical exercise \cite{johannes2007, godin2008analysis, trouvain2015}. Similar effects on speech production have been observed in scenarios eliciting cognitive load \cite{rothkrantz2004voice, johannes2007}. As a matter of fact, paralinguistic studies of stress have historically relied on handcrafted acoustic feature sets based on digital signal processing (DSP). Such feature sets could then be used for classification of cognitive or physical load levels using various classical machine learning approaches, including GMM supervector systems \cite{kua2014unsw}, $i$-vector frameworks \cite{van2014classification}, and ensemble learning \cite{jing2014ensemble}. 

Alternatively, several deep learning frameworks have been proposed for cognitive and physical load detection \cite{gosztolya2014detecting, gallardo2019}. Yet, training such models directly on task load corpora remains a challenging issue due to their small sizes, as larger networks tend to overfit due to their higher model capacity. 
To overcome this issue, several studies have shown that pre-training models on large corpora of speech or audio can produce state-of-the-art performance in various data-constrained speech processing and paralinguistic tasks such as speech emotion recognition~\cite{scheidwasser2021serab}, speaker or language identification~\cite{shor2020, niizumi2021byol, shor2021universal}.

With that in mind, in this paper, we focus on studying the detection of cognitive and physical load from voice recording using data-driven audio representation models. We introduce four novel datasets related to cognitive load with varying task complexities, languages, and protocols, and one dataset related to physical load. We used the datasets to evaluate a novel \textit{hybrid} training protocol for general-purpose speech representation (Figure \ref{fig:diagram}). The proposed approach aims at taking advantage of both self-supervised representation learning and non-trainable DSP-based feature sets. However, unlike existing ensemble approaches, which simply combine outputs from different models, our method incorporates the two models during the training phase to learn a new and unique representation better than the sum of its parts. We employed the proposed model to detect cognitive and physical stress from voice recordings in the above-outlined datasets. We showed that this \textit{hybrid} speech representation generalizes across different datasets and produces better performance than existing DSP-based, as well as DNN-based representation learning methods. Importantly, the proposed hybrid model yielded better results than the ensemble of the methods used for its pre-training.

\setlength{\tabcolsep}{3pt}
\begin{table*}[h!]
\centering
\ninept
\caption{Cognitive/physical load speech datasets summary. ENG: English, FRA: French, CMN: Mandarin Chinese.}
\vspace{-4pt}
\begin{tabular}{llcccl}
\toprule
Dataset & Language & Utterances & Speakers & Duration (h) & Speech data acquisition protocol \\
\midrule 
\textbf{Cog1} & ENG/FRA & 586 & 21 & 1.75 & Count audio stimuli (syllables) while reading scripted text \\
\textbf{Cog2}  & CMN & 800 & 20 & 3.33 & Memorise three \textbf{numeric} digits while reading scripted text\\
\textbf{Cog3}  & CMN & 800 & 20 & 3.33 & Memorise three \textbf{alphanumeric} digits while reading scripted text\\
\textbf{Cog4}  & 9 languages & 349 & 22 & 2.93 & Describe an image while performing four cognitive tasks\\
\textbf{Phys} & ENG/FRA & 280 & 28 & 0.70 & Spontaneous speech before/after VO$_2$ max running test\\
\bottomrule
\end{tabular}
\label{tab:datasets}
\vspace{-9pt}
\end{table*}
\setlength{\tabcolsep}{6pt}

\section{Materials \& Methods}
\label{experiments}
\subsection{Tasks and Datasets}
The speech datasets proposed in this study vary in language, duration, protocol, and more importantly task complexity. A brief description of the datasets is addressed in Table 1. Cognitive load was induced in the speakers through multitasking, based on the idea that cognitive load increases with 
the number of tasks to be performed simultaneously \cite{ryu2005evaluation, chong2018}. On the other hand, physical load was induced by recording speech samples before and after the speaker performed an intensive aerobic physical exercise (VO$_2$ max running test). In each dataset, the recorded speech samples were labelled in accordance with the experimental condition (i.e., with load / without load). 


\noindent\textbf{Cognitive Load 1} (\textbf{Cog1}) - This dataset contains 21 speakers (3 female, 18 male) who participated in their native language (English or French). The experiment comprises six blocks. The first block consisted of a scripted speech task, in which participants read aloud at most three sentences from SIWIS, a multilingual speech and text database~\cite{goldman2016siwis}. The other five blocks included the same scripted speech task, but this time participants were instructed to count spoken syllables mixed in additive Gaussian white noise (AGWN) while reading aloud. Syllables were selected from the Articulation Index dataset \cite{wright2005}. Each block contained five trials lasting 10 seconds each, producing approximately five minutes of speech material per speaker.

\noindent\textbf{Cognitive Load 2} (\textbf{Cog2}) - 
20 Chinese Mandarin speakers (10 female, 10 male) were recruited to perform a similar task to the Cog1 task. The experiment included 40 trials per participant: 20 of them included only reading scripted text from randomly selected articles across various media sources, while the other 20 trials comprised an additional working memory task, as participants had to listen to and memorise three numeric digits while reading. Each trial lasted 15 seconds, giving 10 min of speech material per participant.


\noindent\textbf{Cognitive Load 3} (\textbf{Cog3}) - This task was identical to Cog2, except for the working memory task, as participants were asked to memorize alphanumeric digits (e.g., ``A5" instead of ``5" in Cog2). The dataset comprised 40 15-second trials from 20 speakers, producing 10 min of speech data per speaker.

\noindent\textbf{Cognitive Load 4} (\textbf{Cog4}) - This dataset was collected from 22 subjects (7 female, 15 male). Audio recordings were collected as participants performed an experiment in which cognitive load was induced based on multitasking, with an experimental design inspired by \cite{holm2009estimating}. The experiment was carried out in three blocks, a baseline block lasting four minutes, followed by two multitasking blocks lasting two minutes each. In the baseline block, participants were shown an image on a screen and were asked to describe it in as much detail as possible in their native language. The images shown on the screen were presented in colour and included cityscapes, landscapes, and famous paintings. Every 30 seconds, a new image appeared, and the participants began a new recording. In the multitasking blocks, a high level of cognitive load was applied by asking participants to perform four tasks simultaneously, in addition to describing the image appearing on the screen. The four tasks consisted of an auditory response task, a visual vigilance task, an arithmetic task, and a memory task. The participants were instructed to pay equal attention to all tasks and perform them simultaneously to the best of their ability, while at the same time describing the image on the screen in as much detail as possible. For these blocks, images were selected from commonly used images to study various mental and cognitive impairments through speech (e.g., Cookie-theft picture \cite{goodglass2001bdae}). As in the baseline block, the presented image changed every 30 seconds. Each participant produced a total of eight minutes of speech data.

\noindent\textbf{Physical Load} (\textbf{Phys}) - 28 participants (13 female, 15 male) were recruited to perform a VO$_2$ max running test \cite{leger_indirect_1980}. Before and after the exercise, participants were instructed to speak spontaneously in their native language (English or French) for three minutes about personal topics, e.g., their current work, their favourite physical activities, or their daily routine. Subsequently, 10 utterances were extracted from each speaker, five for each condition (before and after exercise), where each utterance lasted between 5 and 15 seconds.

\subsection{Audio Representation Models}
All cognitive and physical load speech datasets were used as downstream tasks to benchmark different deep learning-based audio representations against well-established handcrafted acoustic feature sets. Importantly, the data-driven models were not fine-tuned directly on the task-specific data, but only pre-trained on independent datasets.



\subsubsection{Handcrafted models}
\noindent\textbf{openSMILE (OS)} - openSMILE~\cite{eyben2010opensmile} is an open-source feature extraction toolkit that generates handcrafted low-level descriptors (LLD) of audio inputs. This toolkit comprises three feature sets obtained by computing functionals on LLD contours. The most prominent of them is ComParE (OS/ComParE) \cite{schuller2013interspeech}, which contains 6373 static features. Another commonly used feature set is eGeMAPS (OS/eGeMAPS) \cite{eyben2015geneva} which includes 88 features. Both approaches are purely DSP-based and require no pre-training on auxiliary datasets.


\subsubsection{Data-driven models}

\noindent\textbf{VGGish} - VGGish~\cite{hershey2017} is a commonly used audio feature extractor derived from the convolutional neural network VGG-16~\cite{Simonyan15}. This model was pre-trained in a supervised fashion to classify audio events from the Youtube-8M dataset~\cite{abu2016}.

\noindent\textbf{YAMNet} - YAMNet~\cite{plakal2020} is a DNN pre-trained to classify events from AudioSet, a large collection of audio samples from YouTube videos~\cite{gemmeke2017audio}. Its feature extractor employs the MobileNetV1 convolutional architecture~\cite{howard2017mobilenets}.

\noindent\textbf{TRILL, layer 19} - TRILL~\cite{shor2020} is a model pre-trained in a self-supervised manner on speech samples from AudioSet. The CNN model based on ResNetish~\cite{hershey2017} was originally designed as a universal speech representation. It uses triplet-loss learning to create an embedding space where temporally adjacent samples are mapped close together. Here, we use the features generated from layer 19, as found by \cite{shor2020} to perform best on non-semantic speech tasks. 

\noindent\textbf{BYOL-A} - Unlike contrastive learning frameworks, Bootstrap Your Own Latent for Audio (BYOL-A)~\cite{niizumi2021byol} generates audio representations using two augmented views of a single audio sample, inspired by the success of BYOL~\cite{grill2020bootstrap} for image representation. To obtain audio representations, the log-mel spectrogram (LMS) of an input audio sample is first fed to a data enhancement module, producing two randomly augmented copies of the input LMS (Figure \ref{fig:diagram}). Subsequently, the augmented spectrograms are respectively passed to an online and a target network. Both networks share a similar architecture, with an encoder and a projector module. However, the online network comprises an additional module called predictor to avoid having collapsed representations \cite{grill2020bootstrap}. The aim of this training paradigm is for the online network to predict the output generated by the target network by minimising a mean squared error (MSE) loss function. BYOL-A has achieved competitive performance in various speech processing tasks, e.g., language and speaker identification.  



\noindent\textbf{BYOL-A Extensions} -  Following the same implementation of BYOL-A\footnote{\url{https://github.com/nttcslab/byol-a}}, we proposed in a previous work BYOL-S\footnote{\url{https://github.com/Neclow/SERAB}}~\cite{scheidwasser2021serab}, BYOL for speech, which was pre-trained on a speech subset of AudioSet. In addition, we modified the default CNN encoder architecture used in BYOL-A (a 5-layer CNN) using a lightweight version of the Convolutional vision Transformer (CvT)\footnote{\url{https://github.com/lucidrains/vit-pytorch}}~\cite{wu2021cvt} to leverage the best of CNNs and Transformers. All proposed encoders produce an embedding of size 2048. The proposed models and associated methods are openly available online
\footnote{\url{https://github.com/GasserElbanna/serab-byols}}.

\subsubsection{Hybrid models}
Considering the effectiveness of handcrafted features for various paralinguistic challenges \cite{schuller2013interspeech, schuller2014interspeech}, we derived a new training protocol for BYOL-S by combining hand-made and data-driven features in a hybrid format. More specifically, we augmented the BYOL framework with a supervision network which generates DSP-based features extracted from the ComParE feature set in openSMILE, as shown in Figure \ref{fig:diagram}. Therefore, in addition to the self-supervised BYOL loss ($\mathcal{L}_{\text{ss}}$), we computed a supervised loss $\mathcal{L}_{\text{sup}}$ between the outputs of the supervised and online networks , set as a MSE. The final loss function was defined as a weighted sum of the self-supervised and supervised losses: 
\begin{equation}
	\mathcal{L}_{\text{hybrid}} = \alpha_{\text{ss}} \mathcal{L}_{\text{ss}} + \alpha_{\text{sup}} \mathcal{L}_{\text{sup}}
\label{equ:loss}
\end{equation}

where $\alpha_{\text{ss}}$ and $\alpha_{\text{sup}}$ weigh the parts of the self-supervised and supervised losses, respectively. 

Given the size of the ComParE feature set (6373), we projected the BYOL-S model outputs, from both online and target networks, to produce 6373-dimensional embeddings to compute the hybrid loss ($\mathcal{L}_{\text{hybrid}}$). However, the output size of the pre-trained encoder module remained unchanged (2048).

\subsection{Evaluation Pipeline}

All models were evaluated on the five cognitive and physical load tasks mentioned above. Audio samples were processed to produce their representation vectors per model.
Each downstream dataset was divided into training and test sets with ratios of 70\% and 30\%, respectively. To ensure a fair and robust evaluation, the partitions were constructed to be speaker-independent (i.e., all utterances from a given speaker belonged to either the training or the test set) and to have a similar gender distribution. Each partition was also standardised (to zero mean and unit variance) to alleviate inter-speaker variability. Subsequently, classification was performed by training and validating linear support-vector machines (SVMs) using five-fold cross-validation. A grid search was applied to optimise the penalty hyperparameter, with values ranging from 10\textsuperscript{-5} to 10\textsuperscript{5}. Using such a simple classifier with linear kernel allows us to place more emphasis on the predictive power of the extracted features. Lastly, model performance was evaluated by computing the unweighted average recall on the test set to compensate for any imbalanced data distributions. All procedures in the evaluation pipeline were implemented using \texttt{scikit-learn}~\cite{scikit-learn}.



\begin{table*}[h!]
  \centering
  \caption{Test
  UAR (\%) on cognitive and physical load datasets. \{\} indicates the concatenation of feature vectors. OS: openSMILE~\cite{eyben2010opensmile}. The best scores are shown in bold. The models are sorted according to the UAR averaged across the five tasks.}
  \vspace{-6pt}
  \begin{tabular}{lccccc@{\hspace{20pt}}ccc}
\toprule
Model & Cog1 & Cog2 & Cog3 & Cog4 & Phys & Average \\
\midrule
VGGish\cite{hershey2017}                                                                                             & 64.0   & 69.6 & 90.0   & 77.1 & 55.0   & 71.1 \\
TRILL, layer 19\cite{shor2020}                                                                                       & 58.8 & 72.9 & 93.3 & 74.0   & 62.5 & 72.3 \\
OS/eGeMAPS\cite{eyben2015geneva}                                                                                     & 62.9 & 65.4 & 93.8 & 72.9 & 67.5 & 72.5 \\
BYOL-A\cite{niizumi2021byol}                                                                                         & 68.0   & 70.4 & 90.0   & 81.3 & 61.3 & 74.2 \\
YAMNet\cite{plakal2020}                                                                                              & 58.1 & 71.7 & 99.2 & 83.3 & 62.5 & 75.0   \\
BYOL-S\cite{scheidwasser2021serab}                                                                                   & 65.3 & 72.9 & 91.7 & 80.2 & 66.3 & 75.3 \\
OS/ComParE\cite{eyben2010opensmile}                                                                                  & 78.7 & 74.6 & \textbf{100.0}  & 79.2 & 52.5 & 77.0   \\
BYOL-S/CvT\cite{scheidwasser2021serab}                                                                               & 79.2 & \textbf{80.0}   & \textbf{100.0}  & 74.0   & 57.5 & 78.1 \\
\midrule
\textbf{\textit{Hybrid \& ensemble models:}} \\
\{BYOL-S\cite{scheidwasser2021serab};   OS/ComParE\cite{eyben2010opensmile}\}     & 76.2 & 77.9 & \textbf{100.0}  & 83.3 & 56.3 & 78.7 \\
\{BYOL-S/CvT \cite{scheidwasser2021serab};   OS/ComParE\cite{eyben2010opensmile}\} & 81.2 & 77.1 & \textbf{100.0}  & 79.2 & 57.5 & 79.0  \\
Hybrid BYOL-S ($\alpha_{\text{ss}} = \alpha_{\text{sup}} = 1$)                                                                                          & 66.5 & 75.4 & 90.8 & \textbf{93.8} & \textbf{71.3} & 79.5 \\
Hybrid BYOL-S/CvT ($\alpha_{\text{ss}} = \alpha_{\text{sup}} = 1$)                                                                                      & 78.4 & 74.6 & 99.2 & 83.3 & 66.3 & 80.3 \\
Hybrid BYOL-S/CvT ($\alpha_{\text{ss}} = 1$, $\alpha_{\text{sup}} = 2$)                                                                                   & \textbf{83.8} & 72.5 & 98.3 & 87.5 & 70.0  & \textbf{82.4} \\
  \bottomrule
  \end{tabular}
  \label{tab:2}
\end{table*}

\begin{table*}[t]
  \centering
  \caption{Impact of hybrid loss weighting (Eq.~\ref{equ:loss}) in \textbf{Hybrid BYOL-S/CvT} pre-training on test UAR (\%).}
  \vspace{-4pt}
  \begin{tabular}{lccccc@{\hspace{20pt}}c}
  \toprule
    $\alpha_{\text{ss}}$ : $\alpha_{\text{sup}}$ (Eq.~\ref{equ:loss}) &
    Cog1 & Cog2 & Cog3 & Cog4 & Phys & Average \\
    \midrule
4 : 1              & 73.5          & 71.3          & 97.5          & 87.5          & 60.0          & 77.9 \\
2 : 1              & 72.2          & 72.5          & \textbf{99.2} & \textbf{92.7} & 55.0          & 78.3  \\
4 : 3              & 76.7          & 70.0          & 98.3          & 88.5          & 62.5          & 79.2          \\
1 : 1              & 78.4          & 74.6          & \textbf{99.2} & 83.3          & 66.3          & 80.3           \\
3 : 4              & 80.2          & \textbf{75.0} & \textbf{99.2} & 86.5          & 65.0          & 81.2           \\
1 : 2              & \textbf{83.8} & 72.5          & 98.3          & 87.5          & \textbf{70.0} & \textbf{82.4}  \\
1 : 4              & 81.4          & 72.5          & 97.9          & 86.5          & 63.8          & 80.4          \\
    \bottomrule
  \end{tabular}
  \label{tab:3}
\end{table*}

  

\section{Results}
\label{results}
Table \ref{tab:2} presents the performance of each model in all five tasks by reporting the unweighted average recall (UAR, in percentages) on an unseen and speaker-independent test set. 
Importantly, using the ComParE handcrafted features exhibited better performance than most deep learning-based models, including recent self-supervised models for speech and audio representation (TRILL, BYOL-A and BYOL-S). Only BYOL-S/CvT achieved better performance, although only slightly better (1.1\% average improvement). This result highlights the effectiveness of DSP-based features in paralinguistic tasks.



On the other hand, hybrid representation models, which aim to take advantage of both data-driven and handcrafted features, yielded significantly better performance. The two proposed hybrid models, one using BYOL-A's CNN encoder \cite{niizumi2021byol} and one with CvT encoding \cite{wu2021cvt}, consistently outperformed their corresponding BYOL-S models. This result suggests that adding the DSP-based supervision to the self-supervised representation learning framework helped to improve its generalization capacity in cognitive/physical load detection tasks.

As the proposed hybrid loss consists of a weighted sum of its self-supervised and supervised counterparts (Eq. \ref{equ:loss}), we reported model performance on the five datasets for different $\alpha_{\text{ss}}$~:~$\alpha_{\text{sup}}$ ratios. Here, we selected the \textbf{Hybrid BYOL-S/CvT} variant, which yielded the best overall performance. The results shown in Table~\ref{tab:3} indicate a steady increase in model performance with increasing weight assigned to the supervised loss ($\alpha_{\text{sup}}$). The 1:2 loss ratio ($\alpha_{\text{ss}}$~:~$\alpha_{\text{sup}}$) yielded the best model for detecting cognitive and physical load in the considered tasks. Increasing the weight of supervised loss further led to a consistent decrease in performance across all tasks.

Finally, we further validated the proposed hybrid framework by examining the results obtained by concatenating ComParE features with either BYOL-S or BYOL-S/CvT embeddings (8421-D feature vector).
In particular, the best hydrid BYOL-S/CvT model yielded a 3.4\% higher UAR than the concatenation of the two representations used in its pre-training. 



\section{Discussion \& Conclusions}
\label{Conclusions}
In this paper, we introduced four novel datasets for large-scale detection of cognitive and physical load from speech. 
We used the datasets to evaluate existing DPS-based and DNN-based speech representation learning methods, and to develop a novel hybrid (supervised / self-supervised) approach.

In spite of the effectiveness of data-driven features, handcrafted acoustic feature sets maintained competitive performance, a result consistent with outcomes of previous paralinguistic challenges \cite{schuller2014interspeech}. To leverage their efficacy, we proposed a novel hybrid approach obtained by combining data-driven features from a BYOL-derived model (BYOL-S/CvT) with handcrafted features extracted from openSMILE~\cite{eyben2010opensmile} during pre-training. This hybrid training protocol outperformed all data-driven and handcrafted feature sets presented herein, including fusions of different individual representations. Unlike model ensemble methods, in the inference stage, the proposed hybrid model involves only a single encoder (i.e., a single forward-pass), thus making it lighter \& faster than conventional approaches, which usually require processing of the input samples through several models. Following the experimentation with hybrid loss weighting, we found that shifting the loss emphasis to the fixed handcrafted features improved the overall model performance. However, increasing this emphasis beyond the optimal loss weighting ratio ($\alpha_{\text{ss}}=1$ \& $\alpha_{\text{sup}}=2$, Eq.~\ref{equ:loss}) led to a poorer performance, which illustrates the synergistic interaction between the two parts of the hybrid framework.

Importantly, the obtained representation differs from a simple concatenation of BYOL-S/CvT and openSMILE features. Indeed, the hybrid training protocol allowed the model to learn a more robust speech representation for detecting cognitive and physical load. The inclusion of an auxiliary loss to update the online network from projected DSP-based features likely helps improve pre-training stability, a known issue in BYOL-derived frameworks \cite{grill2020bootstrap} (usually circumvented by exponential averaging of weight updates). Furthermore, the use of fixed DSP-based features in the representation learning, likely reduces any potential overfitting to the pre-training dataset. As a matter of fact, such a hybrid model also produced highly competitive results within the HEAR benchmark, outperforming several state-of-the-art approaches \cite{elbanna2022}.


The proposed hybrid representation learning approach could be easily incorporated into any existing data-driven audio representation learning methods. Moreover, both components of the proposed framework could be substituted with other (potentially larger/complex)  models. In particular, the fixed DSP-based feature extractor could be replaced with another pre-trained DNN-based representation. In a similar vein to self-distillation approaches \cite{caron2021}, the two learnable representations could be iteratively updated in the pre-training process to produce a potentially more robust
representation. Alternatively, the DSP-based feature extractor could also generate a specific family of acoustic features (e.g., prosodic features) which could be beneficial for building 
more specialized audio representations.



\clearpage
\ninept
\bibliographystyle{IEEEtran}
\bibliography{references}

\end{document}